\documentclass{aac}
\usepackage[numbers]{natbib}
\usepackage{graphicx}

\begin{document}

\title{Carrier-Envelope Phase Controlled Electron Injection into a Laser-Wakefield Accelerator}

\author[aff1]{Jihoon Kim\corref{cor1}}
\author[aff1]{Tianhong Wang}
\author[aff1,aff2]{Vladimir Khudik}
\author[aff1]{Gennady Shvets}

\affil[aff1]{School of Applied and Engineering Physics, Cornell University, Ithaca, NY 14850, USA.}
\affil[aff2]{Department of Physics and Institute for Fusion Studies, The University of Texas at Austin, Austin, TX 78712, USA.}
\corresp[cor1]{jk2628@cornell.edu}

\maketitle

\begin{abstract}
 Single cycle laser pulse propagating inside a plasma causes controllable asymmetric plasma electron expulsion from laser according to laser carrier envelope phase (CEP) and forms an oscillating plasma bubble. Bubble's transverse wakefield is modified, exhibiting periodic modulation. Injection scheme for a laser wakefield accelerator combining a single cycle low frequency laser pulse and a many cycle high frequency laser pulse is proposed. The co-propagating laser pulses form a transversely oscillating wakefield which efficiently traps and accelerates electrons from background plasma.
By tuning the initial CEP of the single cycle laser pulse, injection dynamics can be modified independently of the many cycle pulse, enabling control of electron bunches’ spatial profile.
\end{abstract}

\section{INTRODUCTION}
Plasma-based accelerators offer a new pathway to compact accelerators, supporting unparalleled acceleration gradients in order of hundreds of GV/m. In a laser-wakefield accelerator (LWFA)~\cite{Malka,RMPS}, a plasma cavity is generated via the time-averaged (ponderomotive) pressure of an ultra-intense laser pulse pushing plasma electrons out of its way. The LWFA concept has proven to be highly successful in generating multi-GeV, low-emittance, ultra-short electron bunches at multiple laser facilities around the world \cite{Nakamura, Leemans, Gonsalves}. High-energy electrons generated by LWFAs are promising candidates for various applications, including TeV-scale next-generation lepton colliders~\cite{TeV}, and high brightness radiation and particle sources~\cite{Stark}.

Under most circumstances relevant to laser-plasma accelerators, ponderomotive description is sufficient for describing the response of plasma electrons to laser pulses. Under the ponderomotive approximation, laser pulse polarization and carrier phase play negligible roles~\cite{Mora}. This approximation breaks down for single or few-cycle laser pulses via several mechanisms, including the ionization injection~\cite{chen_jap06} of inner-shell electrons into a plasma bubble that occurs in gaseous plasma targets containing high-$Z$ ions~\cite{lifschitz_malka_njp12}. Another polarization- and phase-dependent mechanism is the creation of an asymmetric plasma "bubble" with an overall displacement in the laser polarization direction. The displacement of a plasma bubble produced by a few-cycle laser pulse propagating through the plasma in the $x-$direction with the group (phase) velocity $v_{\rm g}$ ($v_{\rm ph}$) is determined by the oscillating carrier envelope phase $\Phi_{\rm CEP}(x)$ (CEP). Such CEP oscillations induce a periodic bubble displacement with the same period $v_{\rm g} T_{\rm CEP} \approx \lambda_L v_{\rm g}/\left( v_{\rm ph} -v_{\rm g} \right)$~\cite{kost_cep}. When combined with robust self-injection of plasma electrons into the bubble~\cite{lifschitz_malka_njp12,kost_cep,ma_screp16,CEP_observable}, the jitter of the latter can induce betatron oscillations of the former, resulting in X-ray generation~\cite{CEP_observable}.

While electron injection and acceleration by few-cycle laser pulses has been theoretically and experimentally demonstrated~\cite{CEP_observable,kHz_injection,Veisz_2cycle}, this approach to LWFA has limited potential for high-energy (GeV-scale) electron acceleration. The short pulse length and small pulse energy results in a relatively short acceleration distance \cite{kHz_injection} because of reduced self-focusing~\cite{Sprangle_guide} and rapid diffraction of the laser pulse. Using high-density plasmas to increase self-focusing results in continuous injection, low energy gain, and reduced quality of the accelerated beam~\cite{kHz_injection, Veisz_2cycle}. 

\begin{figure}[t]
    \includegraphics[width=0.4\textwidth]{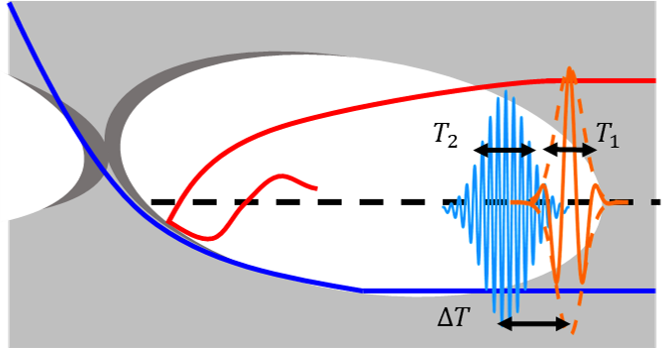}
\caption{Schematic of the LWFA with CEP-based injection. A transversely oscillating plasma bubble driven by a single-cycle injector pulse (orange) and a multi-cycle driver pulse (light blue) periodically traps electrons from ambient plasma. Time-dependent CEP controls bubble center displacement from the laser pulse propagation axis (dashed line), determining which electrons are injected (red line) and which ones pass through (blue). }\label{fig:schematic}
\end{figure}

In this paper, we show that a single-cycle laser pulse can be used as a controllable self-injection mechanism for high-energy acceleration when combined with a multi-cycle higher-intensity laser pulse that serves as the LWFA driver over long distances (see Fig.~\ref{fig:schematic}). We show that this CEP-based Injection (CEPI) mechanism can be realized with moderate-power injector pulses, and is controlled by the CEP of the injector pulse. Below we show the CEPI concept via a single particle model, derive the injection criteria, and validate the results using 3D Particle-In-Cell (PIC) simulations.

\section{Theory and Test-Particle Simulations}
For qualitative understanding of electron injection into an oscillating plasma bubble, we use a simplified model of a positively-charged (devoid of electrons) spherical plasma bubble with radius $R$  propagating with uniform velocity $v_b$~\cite{kostyukov_pop04,kalmykov}. A moving-frame Hamiltonian describing plasma electrons interaction with the bubble is given by $H(\rho,t) = \sqrt{1+(\mathbf{P}+\mathbf{A})^2} - v_b P_x - \phi $~\cite{kalmykov,kost_injection,austin_injection}, where $\mathbf{\rho} = (\xi,y,z-z_{\rm osc})$, $\xi=x-v_b t$, $z_{\rm osc}(t)$ is the transverse coordinate of the oscillating bubble center, $\mathbf{P}$ is the canonical momentum, and $\mathbf{A}$ ($\phi$) are the vector (scalar) potentials. For simplicity, the time, length, potential, and electron momentum variables are normalized to $\omega_p^{-1}$, $k_p^{-1} = c/\omega_p$, $m_e c^2/|e|$, and $m_e c$, respectively. Under the $A_x = -\phi = \Phi/2$  gauge, we further assume that $\Phi=(\rho^2-R^2)/4$ inside and $\Phi=0$ outside the bubble~\cite{kost_injection}. Transverse oscillations $z_{\rm osc}(t) \equiv z_{0}\cos(\omega_{\rm CEP} t + \phi_{\rm CEP})$ of the plasma bubble excited by the injector laser pulse introduces time dependence into the Hamiltonian. Here $z_{0}$ is the maximum bubble oscillation amplitude, and $\phi_{\rm CEP}$ has the meaning of the CEP phase at the time of the electron's entrance into the bubble. 


The electron equations of motion in the $x-z$ plane can be derived from $H(\rho,t)$:
\begin{eqnarray}
&&\frac{d p_x}{dt} =-\frac{(1+v_b)\xi}{4}+ \frac{v_z\tilde{z}(t)}{4}-\frac{\tilde{z}(t)\dot{z}_{osc}(t)}{4}, \label{eq:EqM_N1}\\
&&\frac{d p_{z}}{dt} = -\frac{(v_x+1)\tilde{z}(t)}{4},\label{eq:EqM_N2}
\end{eqnarray}
where the relativistic electron momenta $p_{x,z}$ are related to the corresponding electron displacements according to $d \xi/dt = p_x/\gamma-v_b$ and $d z/dt = p_z/\gamma$. Finally, $\tilde{z}(t)=z-z_{osc}$  defines electron displacement from the oscillating bubble center.

Electrons have $H=1$ prior to their interaction with the bubble, and for robust injection to occur, Hamiltonian must evolve such that $H<0$ ~\cite{kalmykov,austin_injection}. Without time-dependent terms $z_{\rm osc}(t)$ and $\dot{z}_{\rm osc} \equiv d z_{\rm osc}/dt$, the Hamiltonian is conserved ($dH/dt=0$), and electrons cannot get trapped/accelerated by the plasma bubble except for large radius of bubble\cite{kost_injection}.  Bubble oscillations cause the Hamiltonian to evolve, and enable the trapping of plasma electrons:$
    \frac{\partial H}{\partial t} = \frac{dH}{dt} =\dot{p_z}(t)\dot{z}_{osc}(t),
$
where the condition for trapping is given by integral over electron trajectory: $\Delta H =\int  \dot{p_z}(t)\dot{z}_{osc}(t) dt < -1$. 

 Assuming that the electron passage time through the bubble, $T_{\rm pass}\sim R$, is much shorter than $T_{\rm CEP}$, the Hamiltonian increment can be approximated as $\Delta H^{(1)} \approx - z_{0}\omega_{\rm CEP} \sin(\phi_{\rm CEP}) \Delta p^{(0)}$, where $\Delta p^{(0)}$ is the zeroth-order ($z_{\rm osc}=0$) transverse momentum increment of an electron passing through the bubble. For an electron entering the bubble at its edge at $z=\pm R$ and pulled into the interior,  the maximum transverse momentum gain is $|p_z^{\rm max}| \approx 0.14 R^2$ in the limit of $v_b=c$ \cite{kost_injection}. Therefore, trapping condition for an electron entering the bubble's edge at the right phase ($\phi_{\rm CEP}=\pi/2$) can be estimated by $z_0 >7/(\omega_{\rm CEP}R^2)$. In physical units, the injection criterion can be expressed as $z_0\sim k_p^{-1}\sqrt{n_{\rm crit}/n_p}\left(k_p R \right)^{-2}$ .

\begin{figure}[t]
    \includegraphics[width=1\textwidth]{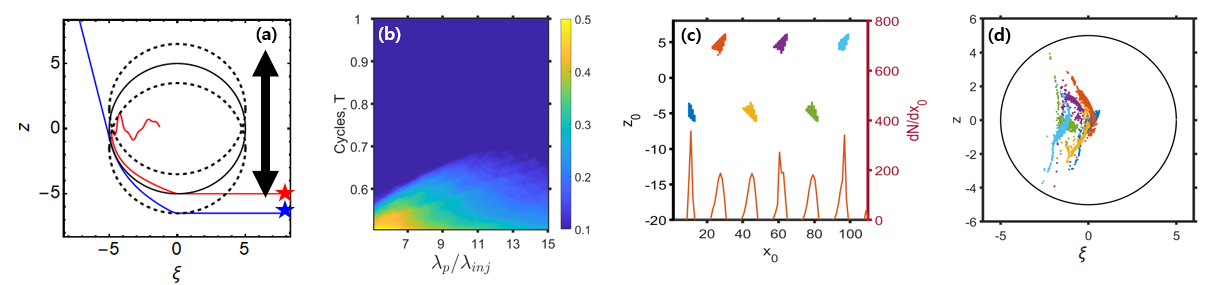}
\caption{
Test particles results. (a) Trajectories in the $\xi-z$ plane and moving reference frame: $\phi_{\rm CEP}=\pi/2$ (red) and $\phi_{\rm CEP}=\pi$ (blue). Bubble boundaries: unperturbed (black solid line) and maximally-displaced (black dashed line). (b) Parameter scan over laser wavelength and cycles of average electron transverse momentum $|\langle p_{z}\rangle|$ after passing through a single-cycle pulse and a quiescent bubble. (c-d) Injected electrons from distinct initial locations are color-coded based on their initial location $x_0$ and plotted at $t=300$ (c) in the initial conditions plane $(x_0,z_0)$ and (d) inside the bubble (black circle). Red line in (c): density of injected electrons vs initial position.}\label{fig:trajectory}
\end{figure}

The injection process is visualized in Fig.~\ref{fig:trajectory} (a), where the results of solving equations of motion given by Eqs.~(\ref{eq:EqM_N1}) and (\ref{eq:EqM_N2}) are shown for two initially quiescent electrons. The first (red star) electron enters a transversely oscillating bubble with $\phi_{\rm CEP}=\pi/2$, while the second (blue star) one is delayed in time so as to enter the bubble with $\phi_{\rm CEP}=\pi$.  The electrons' initial transverse position is chosen such that electrons enter the bubble at its lower edge ($\tilde{z}(t_0)=-R,\xi=0$), which is typically the optimal position for electron injection into a plasma bubble~\cite{kalmykov,kost_injection}. The chosen normalized bubble parameters ($R=5$, $z_0/R=0.3$, $T_{\rm CEP}=35$, $\gamma_b= 5$) approximately correspond to a realistic particle-in-cell (PIC) plasma simulation presented later. According to Fig.~\ref{fig:trajectory}(a), only the first electron is injected into the bubble, while the second one passes through. This phase-dependent  behavior can be understood via tracking the evolution of particle's Hamiltonian: the trapped electron's Hamiltonian decreases below zero near $t=15$, fulfilling the sufficient trapping condition. In contrast, the passing electron's Hamiltonian increases and remains positive. In order to be injected into the bubble, an electron must enter it at the appropriate oscillation phase $\phi_{\rm CEP}$ of its center. Since these transverse oscillations are periodic, injection events also happen periodically (twice per period). Specifically, if a bubble can trap electron at $(x_1,z_1)$ for $\phi_{\rm CEP1}$, then  it will also trap a "partner" electron for $\phi_{\rm CEP2} = \phi_{\rm CEP1} + \pi$ at $(x_2=x_1 + v_b T_{\rm CEP}/2,z_2=-z_1)$. 

 In practice, asymmetry of bubble is generated by a single-cycle kicker pulse: to show that such pulse can generate an asymmetric plasma flow and understand dependency on pulse parameters, we ran a test-particle parameter scan where a swarm of particles pass through a combined field of a quiescent bubble and an injector pulse. By measuring the mean transverse momenta of the particles after they pass the combined fields, we quantify the degree of asymmetry. Bubble size and injector pulse spot size is fixed, while the cycle number and ratio between injector pulse wavelength and plasma wavelength were varied. The two parameters are chosen order to take advantage of the nonponderomotive scaling law: $\delta p_z \propto U_p^{3/2} \lambda_{\rm inj}^2/\sigma_{\rm inj}^2 \sin(\phi_{\rm CEP})$~\cite{kost_cep}. Our numerical computation shows [see Fig.~\ref{fig:trajectory}(b)] that there is greater asymmetry in transverse momentum when (1) injector pulse laser wavelength is longer and (2) injector pulse  laser cycle number is smaller, in qualitative agreement with the non-ponderomotive scaling. Combined with our previous injection criterion, this shows that using a long-wavelength injector is advantageous.

To model electron trapping from background plasma, we simulate the interaction of the same oscillating plasma bubble with a swarm of initially resting electrons that are seeded into a three-dimensional (3D)  volume of initial positions $(x_0,y_0,z_0)$ defined by the $R<x_0<100$, $-6.5 < y_0,z_0 < 6.5$ range. The equations of motion are solved for the electrons entering the bubble during the $0<t<300$ time interval. We observe that electron injection is  periodic in time $t=x_0/v_b$, with spacing equal to the half-period of the bubble centroid oscillation $\Delta t = T_{\rm CEP}/2$, as can be seen in Fig.~\ref{fig:trajectory}(c). The injections occur in short bursts at times corresponding to the optimal bubble center oscillation phase $\phi_{\rm CEP}=\pm \pi/2$ at the electron entrance into the bubble. By color-coding the electrons by their injection time and plotting them in the $(t, z_0)$ space, we show that  electrons are injected from alternating initial locations $z_0 = \pm R$. This shows that the particles situated near the bubble's edge are the easiest to trap. This can be understood from the scaling of $|\Delta H| \propto \bar{\rho}^2$, where $\bar{\rho}$ is the impact parameter of an electron entering the bubble. Therefore, those electrons ``grazing" the bubble's edge ($\bar{\rho} \approx R$) have the largest change in their Hamiltonian, making them the best candidates for satisfying the injection criterion.

Time delay between separate injection affects the longitudinal structure of the trapped/accelerated particle beam. As the electrons are accelerated at the back of the bubble, they advance forward through the bubble. Therefore, longitudinal spacing between micro-bunches injected at different times is given by $\Delta \xi \approx (c-v_b) \Delta t  \approx T_{\rm CEP}/(4 \gamma _b^2)$, where $\gamma_b$ is the relativistic factor of the bubble, and $v_b/c \approx 1-1/2\gamma_b^2$ for $\gamma_b \gg 1$. Several micro-bunches are observed in Fig.~\ref{fig:trajectory}(d).

\begin{figure}[t]
\centering
    \includegraphics[width=1\textwidth]{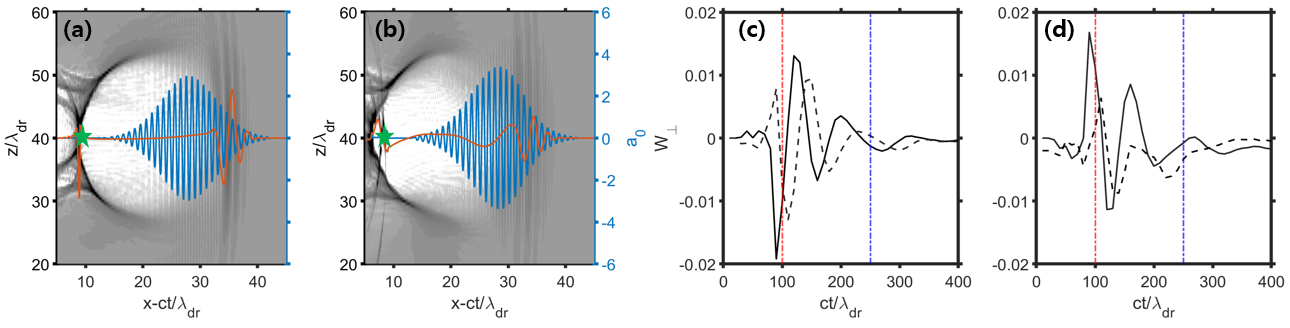}
\caption{Evolution of the laser pulses and plasma wakes. (a-b) Bubble in x-z plane and on axis laser pulse of the driver (blue line) and  $\lambda_{\rm inj}=2.4\,\mu m $ kicker (red line) for  injector pulse $\phi_{\rm CEP}=0$ at  (a) x=0.08 mm (b) x=0.2 mm.  Green star: $\zeta$-location where the  wakefields are plotted in (c-d).
(c-d) On-axis transverse wakes at $\zeta \equiv (x-ct)/\lambda = 9$ (c) for the $\lambda_{\rm inj}=2.4\,\mu m $ injector pulse CEPs  $\phi_{\rm CEP}=0$ (black solid line) and $\phi_{\rm CEP}=\pi/2$ (black dashed line) (d) for the  injector pulse CEPs  $\phi_{\rm CEP}=\pi$ , $\lambda_{\rm inj}=2.4\,\mu m $ (black solid line) and $\lambda_{\rm inj}^{(2)}=1.2\,\mu m $ (black dashed line).    Red (blue) dotted-dashed lines correspond to propagation distances $x_1=0.08$ mm ($x_2=0.2$ mm).
 Plasma wakes and laser electric field strength scale: $E_0=e/mc\omega_{\rm dr}$. Plasma density: $n_p=4.4\times 10^{18}$ cm$^{-3}$, laser parameters: see Table I.}\label{fig:evolution}
\end{figure}\

\section{PIC Simulation Results}

To model complex physical effects accompanying nonlinear interactions between the two laser pulses and the plasma, we use a 3D particle-in-cell (PIC) code VLPL \cite{Pukhov_code}. As shown in Fig \ref{fig:schematic}, two linearly polarized laser pulses are launched into a tenuous plasma with a plateau density $n_p=4.4\times 10^{18}$ cm$^{-3}$ prefaced by a linear density ramp with length $L=0.03\, {\rm mm}$. A multi-cycle short-wavelength driver pulse with the wavelength $\lambda_{\rm dr}$ trails a single-cycle long-wavelength injector pulse with the wavelength $\lambda_{\rm inj}$, with the delay time $\Delta T = 21\, {\rm fs}$ optimized so as to place the injected electrons close to the rear of the plasma bubble formed by the driver pulse. Pulses' parameters are summarized in Table \ref{table:1}. Note that the injector pulse energy $U_{\rm inj} \sim 20\, {\rm mJ}$ in this example is a small fraction of the driver pulse energy $U_{\rm dr} \sim 580\,{\rm mJ}$ because of its low power and short duration.

\begin{table}[h!]
\centering
 \begin{tabular}{c c c c}
 \hline
  Laser pulse& Driver & Injector & Injector 2  \\
 \hline
  Polarization & y & z &z \\  \hline
  Wavelength& $\lambda_{\rm dr} = 0.8\,{\rm \mu m}$ & $\lambda_{\rm inj} = 2.4\,{\rm \mu m}$ & $\lambda_{\rm inj}^{(2)} = 1.2\,{\rm \mu m}$  \\
 \hline
 Duration & $T_{\rm dr}=19\, {\rm fs}$ & $T_{\rm inj}= 5\, {\rm fs}  $ & $T_{\rm inj}^{(2)}= 2.5\, {\rm fs}$\\
 \hline
 Spot Size & $\sigma_{\rm dr} = 10 \,{\rm \mu m}$ &  $\sigma_{\rm inj} = 8\, {\rm \mu m}$ & $\sigma_{\rm inj}^{(2)} = 8\, {\rm \mu m}$\\
 \hline
 Peak power & $P_{\rm dr}= 31\, {\rm TW}$ & $P_{\rm inj}= 3.4\, {\rm TW}$ & $P_{\rm inj}^{(2)}= 13.2\, {\rm TW}$  \\
[1ex]\hline
\end{tabular}
\caption{Pulses' Parameters}
\label{table:1}
\end{table}

During the early stage of co-propagation of the driver and injector pulses [see Fig.~\ref{fig:evolution}(a)], the former produces the bubble while the latter induces its transverse centroid oscillations in the direction of the injector polarization. Bubble oscillations are manifested as a transverse on-axis field wake $W_{\perp}\equiv E_z + B_y$ shown in Fig.~\ref{fig:evolution}(c), where $\left( E_z, B_y \right)(\zeta,z=y=0)$ are the transverse electric/magnetic wakefields of the bubble. During the co-propagation phase lasting over a distance $x\approx 0.2\, {\rm mm}$, the transverse wake oscillates with a spatial period $cT_{\rm CEP} \approx 70 \lambda_{\rm dr}$ that can be estimated from
$  cT_{\rm CEP} \approx 2 \lambda_{\rm inj} \left( \frac{1}{\gamma_b^2} + \frac{n_p}{n_{\rm crit}(\lambda_{\rm inj})} \right)^{-1}$
where $n_{\rm crit}(\lambda)=\pi m_e c^2/(e^2\lambda^2)$ is the critical density corresponding to wavelength $\lambda$. The wakefield phase is easily controlled by the injector CEP as shown in Fig.~ \ref{fig:evolution}(c), where the transverse wakes produced by the injector pulses with two different initial values of $\phi_{\rm CEP}=0,\pi/2$ are phase-shifted by $90^{\circ}$ with respect to each other. On the other hand, the magnitude and phase of the longitudinal wake $W_{\rm parallel}\equiv E_x$ are CEP-independent. 

The difference of our two-pulse scenario from the previously proposed  scheme utilizing a single CEP-controlled injector pulse for injection/acceleration is that the driver pulse outruns the injector pulse as shown in Fig.~\ref{fig:evolution}(b). Rapid depletion of the injection pulse is owing to its much stronger interaction with plasma due to its longer wavelength, as expressed by the $n_p/n_{\rm crit}(\lambda_{\rm inj}) \gg n_p/n_{\rm crit}(\lambda_{\rm dr})$ relationship. Indeed, the field strength of the injection pulse significantly decays after $L_{\rm inj}^{\rm depl}= 250 \lambda_{\rm dr}$ while the driver pulse stays the same as shown in Fig.~\ref{fig:evolution}(d). The injection pulse depletion is mirrored in the decay of the transverse plasma wake $W_{\perp}$ over the same depletion distance $L_{\rm inj}^{\rm depl}$ plotted in Fig.~\ref{fig:evolution}(c). Therefore, we expect the injection to stop after $x \approx L_{\rm inj}^{\rm depl}$ and stable acceleration of the trapped electrons to proceed for a much longer distance, leading to stable high-energy acceleration.

\begin{figure}[h]
\centering
    \includegraphics[width=1\textwidth]{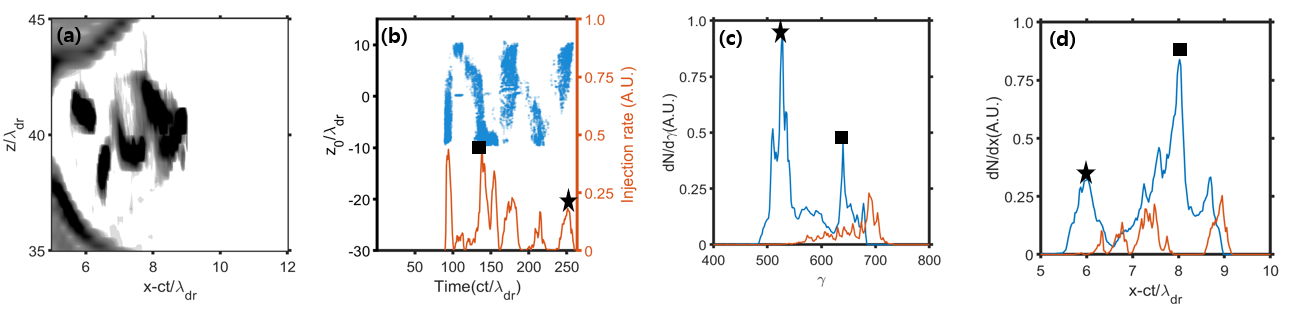}
\caption{Electron injection into an oscillating bubble. (a) Densities of ambient plasma and injected electrons in the $x-z$ plane at $x=0.32 \,{\rm mm}$. (b) Electron injection rate (solid line) and transverse locations (blue dots) of the injected electrons as a function of time. (c) Energy spectrum and (d) Injected electron current at $x=L_{\rm acc} = 1.5\, {\rm mm}$ for different injector pulse wavelengths: $\lambda_{\rm inj}=2.4\,{\rm \mu m}$ (blue line) and $\lambda_{\rm inj}=1.2\,{\rm \mu m}$ (red line). Star and square labels in (b-d) label the corresponding injection times (b), energy spectra (c), and positions inside the bubble (d).}\label{fig:injected}
\end{figure}
 
As shown in test-particle simulations (see Fig.~\ref{fig:trajectory}), periodic oscillations of the plasma bubble's centroid induce periodic injections of electrons into the bubble. This is observed in the PIC simulations and plotted in Fig.~\ref{fig:injected}(a): multiple micro-bunches are trapped and accelerated inside the plasma bubble after the extinction of the injection pulse, but significantly before the depletion of the driver pulse. Electrons are periodically injected from the plasma into the bubble as shown in Fig.~\ref{fig:injected}(b) with a periodic spacing of $T_{\rm CEP}/2 \approx 35 \lambda/c$ corresponding to a half-period of the centroid oscillation. Injection terminates after $x \approx L_{\rm inj}^{\rm depl}$ when the transverse oscillation amplitude decreases to zero as shown in Fig.~\ref{fig:evolution}(c). The total injected charge is calculated to be $Q_1\approx 93\,{\rm pC}$, and distributed over several micro-bunches shown in Fig.~\ref{fig:evolution}(a). We have also found that the injector pulse CEP directly controls the transverse distribution of the electrons inside the plasma bubble: the injected bunches corresponding to $\phi_{\rm CEP}^{\prime} = \phi_{\rm CEP} + \pi$ are mirror images of each other, i.e. $z^{\prime} = -z$ and $\zeta^{\prime} = \zeta$ for the corresponding electrons. Therefore, the CEP phase may be extracted by observing the resulting deflection angle from one of the resulting bunches as they exit the plasma.

Multiple injection events occurring during the co-propagation stage of the two pulses are reflected in the complex electron spectrum after the propagation distance $L_{\rm acc}\approx 1.5\, {\rm mm}$ shown in Fig.~\ref{fig:injected}(c). Correspondingly, the beam current plotted in Fig.~\ref{fig:injected}(d) exhibits multiple peaks separated by $\Delta \xi \approx \Delta T_{\rm cep}/4 \gamma_{bb}^2 \approx 0.66 \lambda_{\rm dr}$, as was explained in the context of the test-particle simulation shown in Fig.~\ref{fig:trajectory} (d). The two prominent quasi-monoenergetic peaks at $\gamma_1 \approx 520$ (star) and $\gamma_2 \approx 670$ (square) correspond to the injection times $ct_1\approx 250 \lambda_{\rm dr}$ and $ct_2\approx 135\lambda_{\rm dr}$, respectively [see Fig.~\ref{fig:injected}(b)]. Electrons injected at earlier time $t_2 < t_1$ reach higher energy because they are positioned ahead of the $t_1$ injection inside the bubble as shown in Fig.~\ref{fig:injected}(d). Therefore, they partially deplete the wakefield, resulting in a smaller acceleration gradient for the later injections. 

The advantage of the presented two-pulse scheme involving a long-wavelength single-cycle injector pulse was verified by simulating the following two scenarios: (i) no injector pulse, (ii) short-wavelength ($\lambda_{\rm inj}^{(2)} = \lambda_{\rm inj}/2$) injector pulse with $P_{\rm inj}^{(2)} = 4P_{\rm inj} = 13.2\, {\rm TW}$ and $\tau = 2.5\,{\rm fs}$. While the injector pulse alone can also inject/accelerate electrons, it cannot sustain a stable accelerating bubble over a significant distance, resulting in the energy gain of less than $10$ MeV. Scenario (i) only yields a small accelerated charge $q\approx 1\, {\rm pC}$. 
The driver pulse cannot efficiently inject electrons into the bubble due to its slowly-evolving nature. Scenario (ii), designed so as to preserve the ponderomotive potential $U_p \propto P_{\rm inj}\lambda_{\rm inj}^2/\sigma_{\rm inj}^2$ of the injector pulse while reducing its wavelength, also results in an inefficient charge injection. The corresponding current profile is indicated by a red line in Fig.~\ref{fig:injected}(d), and the total injected charge $Q_2\approx 21\,{\rm pC} \ll Q_1$ [see Fig.~\ref{fig:injected}(c)]. 
Qualitatively, this occurs because the CEP-dependent transverse momentum $\delta p_z$ imparted to a plasma electron by a single-cycle injection pulse has a non-ponderomotive scaling that favors longer wavelengths. The smaller asymmetry in transverse momentum is evidenced by a smaller $W_\perp$ amplitude compared to $2.4\,\mu$m case, as seen in Fig. ~\ref{fig:evolution} (d).

In conclusion, we propose and theoretically demonstrate a two-pulse CEP-dependent scheme for laser injection and acceleration of electrons from a preformed plasma. By combining a single-cycle long-wavelength laser pulse for rapid electron injection and a multi-cycle short-wavelength driver pulse providing long-distance acceleration of the injected electrons, we demonstrate that a sequence of equally spaced ultra-short (sub-femtosecond) high-current (tens of kA) electron micro-bunches with ultra-relativistic energies (hundreds of MeVs) can be generated. Direct control of the betatron oscillation phase of each micro-bunch is provided by the initial CEP of the injector pulse. We envision that such beams can be used as compact sources of X-ray radiation.

\section{ACKNOWLEDGMENTS}
This work was supported by the DOE Grant No. DESC0019431. The authors thank the Texas Advanced Computing Center (TACC) at The University of Texas at Austin for providing the HPC resources. 


\nocite{*}
\bibliographystyle{aac}%
\bibliography{aac2020_latex}%

\begin{thebibliography}{21}%
\makeatletter
\providecommand \@ifxundefined [1]{%
 \@ifx{#1\undefined}
}%
\providecommand \@ifnum [1]{%
 \ifnum #1\expandafter \@firstoftwo
 \else \expandafter \@secondoftwo
 \fi
}%
\providecommand \@ifx [1]{%
 \ifx #1\expandafter \@firstoftwo
 \else \expandafter \@secondoftwo
 \fi
}%
\providecommand \natexlab [1]{#1}%
\providecommand \enquote  [1]{``#1''}%
\providecommand \bibnamefont  [1]{#1}%
\providecommand \bibfnamefont [1]{#1}%
\providecommand \citenamefont [1]{#1}%
\providecommand \href@noop [0]{\@secondoftwo}%
\providecommand \href [0]{\begingroup \@sanitize@url \@href}%
\providecommand \@href[1]{\@@startlink{#1}\@@href}%
\providecommand \@@href[1]{\endgroup#1\@@endlink}%
\providecommand \@sanitize@url [0]{\catcode `\$12\catcode `\&12\catcode
  `\#12\catcode `\^12\catcode `\_12\catcode `\%12\relax}%
\providecommand \@@startlink[1]{}%
\providecommand \@@endlink[0]{}%
\providecommand \url  [0]{\begingroup\@sanitize@url \@url }%
\providecommand \@url [1]{\endgroup\@href {#1}{\urlprefix }}%
\providecommand \urlprefix  [0]{URL }%
\providecommand \Eprint [0]{\href }%
\providecommand \doibase [0]{http://dx.doi.org/}%
\providecommand \selectlanguage [0]{\@gobble}%
\providecommand \bibinfo  [0]{\@secondoftwo}%
\providecommand \bibfield  [0]{\@secondoftwo}%
\providecommand \translation [1]{[#1]}%
\providecommand \BibitemOpen [0]{}%
\providecommand \bibitemStop [0]{}%
\providecommand \bibitemNoStop [0]{.\EOS\space}%
\providecommand \EOS [0]{\spacefactor3000\relax}%
\providecommand \BibitemShut  [1]{\csname bibitem#1\endcsname}%
\let\auto@bib@innerbib\@empty
\bibitem [{\citenamefont {Malka}\ \emph {et~al.}(2008)\citenamefont {Malka},
  \citenamefont {Faure}, \citenamefont {Gauduel}, \citenamefont {Lefebvre},
  \citenamefont {Rousse},\ and\ \citenamefont {Phuoc}}]{Malka}%
  \BibitemOpen
  \bibfield  {author} {\bibinfo {author} {\bibfnamefont {V.}~\bibnamefont
  {Malka}}, \bibinfo {author} {\bibfnamefont {J.}~\bibnamefont {Faure}},
  \bibinfo {author} {\bibfnamefont {Y.~A.}\ \bibnamefont {Gauduel}}, \bibinfo
  {author} {\bibfnamefont {E.}~\bibnamefont {Lefebvre}}, \bibinfo {author}
  {\bibfnamefont {A.}~\bibnamefont {Rousse}}, \ and\ \bibinfo {author}
  {\bibfnamefont {K.~T.}\ \bibnamefont {Phuoc}},\ }\href@noop {} {\bibfield
  {journal} {\bibinfo  {journal} {Nat. Phys.}\ }\textbf {\bibinfo {volume}
  {4}},\ \unskip\ \bibinfo {pages} {447--453} (\bibinfo {year}
  {2008})}\BibitemShut {NoStop}%
\bibitem [{\citenamefont {Esarey}, \citenamefont {Schroeder},\ and\
  \citenamefont {Leemans}(2009)}]{RMPS}%
  \BibitemOpen
  \bibfield  {author} {\bibinfo {author} {\bibfnamefont {E.}~\bibnamefont
  {Esarey}}, \bibinfo {author} {\bibfnamefont {C.~B.}\ \bibnamefont
  {Schroeder}}, \ and\ \bibinfo {author} {\bibfnamefont {W.~P.}\ \bibnamefont
  {Leemans}},\ }\href@noop {} {\bibfield  {journal} {\bibinfo  {journal}
  {Rev.Mod.Phys}\ }\textbf {\bibinfo {volume} {81}},\ p.\ \bibinfo {pages}
  {1229} (\bibinfo {year} {2009})}\BibitemShut {NoStop}%
\bibitem [{\citenamefont {Nakamura}\ \emph {et~al.}(2007)\citenamefont
  {Nakamura}, \citenamefont {Nagler}, \citenamefont {T´oth}, \citenamefont
  {Geddes}, \citenamefont {Schroeder}, \citenamefont {Esarey},\ and\
  \citenamefont {Hooker}}]{Nakamura}%
  \BibitemOpen
  \bibfield  {author} {\bibinfo {author} {\bibfnamefont {K.}~\bibnamefont
  {Nakamura}}, \bibinfo {author} {\bibfnamefont {B.}~\bibnamefont {Nagler}},
  \bibinfo {author} {\bibfnamefont {C.}~\bibnamefont {T´oth}}, \bibinfo
  {author} {\bibfnamefont {C.~G.~R.}\ \bibnamefont {Geddes}}, \bibinfo {author}
  {\bibfnamefont {C.~B.}\ \bibnamefont {Schroeder}}, \bibinfo {author}
  {\bibfnamefont {E.}~\bibnamefont {Esarey}}, \ and\ \bibinfo {author}
  {\bibfnamefont {S.~M.}\ \bibnamefont {Hooker}},\ }\href@noop {} {\bibfield
  {journal} {\bibinfo  {journal} {Phys. Plasmas}\ }\textbf {\bibinfo {volume}
  {14}},\ p.\ \bibinfo {pages} {056708} (\bibinfo {year} {2007})}\BibitemShut
  {NoStop}%
\bibitem [{\citenamefont {Leemans}\ \emph {et~al.}(2014)\citenamefont
  {Leemans}, \citenamefont {Gonsalves}, \citenamefont {Mao}, \citenamefont
  {Nakamura}, \citenamefont {Benedetti}, \citenamefont {Schroeder},
  \citenamefont {T\'oth}, \citenamefont {Daniels}, \citenamefont
  {Mittelberger}, \citenamefont {Bulanov}, \citenamefont {Vay}, \citenamefont
  {Geddes},\ and\ \citenamefont {Esarey}}]{Leemans}%
  \BibitemOpen
  \bibfield  {author} {\bibinfo {author} {\bibfnamefont {W.~P.}\ \bibnamefont
  {Leemans}}, \bibinfo {author} {\bibfnamefont {A.~J.}\ \bibnamefont
  {Gonsalves}}, \bibinfo {author} {\bibfnamefont {H.-S.}\ \bibnamefont {Mao}},
  \bibinfo {author} {\bibfnamefont {K.}~\bibnamefont {Nakamura}}, \bibinfo
  {author} {\bibfnamefont {C.}~\bibnamefont {Benedetti}}, \bibinfo {author}
  {\bibfnamefont {C.~B.}\ \bibnamefont {Schroeder}}, \bibinfo {author}
  {\bibfnamefont {C.}~\bibnamefont {T\'oth}}, \bibinfo {author} {\bibfnamefont
  {J.}~\bibnamefont {Daniels}}, \bibinfo {author} {\bibfnamefont {D.~E.}\
  \bibnamefont {Mittelberger}}, \bibinfo {author} {\bibfnamefont {S.~S.}\
  \bibnamefont {Bulanov}}, \bibinfo {author} {\bibfnamefont {J.-L.}\
  \bibnamefont {Vay}}, \bibinfo {author} {\bibfnamefont {C.~G.~R.}\
  \bibnamefont {Geddes}}, \ and\ \bibinfo {author} {\bibfnamefont
  {E.}~\bibnamefont {Esarey}},\ }\href {\doibase
  10.1103/PhysRevLett.113.245002} {\bibfield  {journal} {\bibinfo  {journal}
  {Phys. Rev. Lett.}\ }\textbf {\bibinfo {volume} {113}},\ p.\ \bibinfo {pages}
  {245002} (\bibinfo {year} {2014})}\BibitemShut {NoStop}%
\bibitem [{\citenamefont {Gonsalves}\ \emph {et~al.}(2019)\citenamefont
  {Gonsalves}, \citenamefont {Nakamura}, \citenamefont {Daniels}, \citenamefont
  {Benedetti}, \citenamefont {Pieronek}, \citenamefont {de~Raadt},
  \citenamefont {Steinke}, \citenamefont {Bin}, \citenamefont {Bulanov},
  \citenamefont {van Tilborg}, \citenamefont {Geddes}, \citenamefont
  {Schroeder}, \citenamefont {T\'oth}, \citenamefont {Esarey}, \citenamefont
  {Swanson}, \citenamefont {Fan-Chiang}, \citenamefont {Bagdasarov},
  \citenamefont {Bobrova}, \citenamefont {Gasilov}, \citenamefont {Korn},
  \citenamefont {Sasorov},\ and\ \citenamefont {Leemans}}]{Gonsalves}%
  \BibitemOpen
  \bibfield  {author} {\bibinfo {author} {\bibfnamefont {A.~J.}\ \bibnamefont
  {Gonsalves}}, \bibinfo {author} {\bibfnamefont {K.}~\bibnamefont {Nakamura}},
  \bibinfo {author} {\bibfnamefont {J.}~\bibnamefont {Daniels}}, \bibinfo
  {author} {\bibfnamefont {C.}~\bibnamefont {Benedetti}}, \bibinfo {author}
  {\bibfnamefont {C.}~\bibnamefont {Pieronek}}, \bibinfo {author}
  {\bibfnamefont {T.~C.~H.}\ \bibnamefont {de~Raadt}}, \bibinfo {author}
  {\bibfnamefont {S.}~\bibnamefont {Steinke}}, \bibinfo {author} {\bibfnamefont
  {J.~H.}\ \bibnamefont {Bin}}, \bibinfo {author} {\bibfnamefont {S.~S.}\
  \bibnamefont {Bulanov}}, \bibinfo {author} {\bibfnamefont {J.}~\bibnamefont
  {van Tilborg}}, \bibinfo {author} {\bibfnamefont {C.~G.~R.}\ \bibnamefont
  {Geddes}}, \bibinfo {author} {\bibfnamefont {C.~B.}\ \bibnamefont
  {Schroeder}}, \bibinfo {author} {\bibfnamefont {C.}~\bibnamefont {T\'oth}},
  \bibinfo {author} {\bibfnamefont {E.}~\bibnamefont {Esarey}}, \bibinfo
  {author} {\bibfnamefont {K.}~\bibnamefont {Swanson}}, \bibinfo {author}
  {\bibfnamefont {L.}~\bibnamefont {Fan-Chiang}}, \bibinfo {author}
  {\bibfnamefont {G.}~\bibnamefont {Bagdasarov}}, \bibinfo {author}
  {\bibfnamefont {N.}~\bibnamefont {Bobrova}}, \bibinfo {author} {\bibfnamefont
  {V.}~\bibnamefont {Gasilov}}, \bibinfo {author} {\bibfnamefont
  {G.}~\bibnamefont {Korn}}, \bibinfo {author} {\bibfnamefont {P.}~\bibnamefont
  {Sasorov}}, \ and\ \bibinfo {author} {\bibfnamefont {W.~P.}\ \bibnamefont
  {Leemans}},\ }\href {\doibase 10.1103/PhysRevLett.122.084801} {\bibfield
  {journal} {\bibinfo  {journal} {Phys. Rev. Lett.}\ }\textbf {\bibinfo
  {volume} {122}},\ p.\ \bibinfo {pages} {084801} (\bibinfo {year}
  {2019})}\BibitemShut {NoStop}%
\bibitem [{\citenamefont {Schroeder}\ \emph {et~al.}(2010)\citenamefont
  {Schroeder}, \citenamefont {Esarey}, \citenamefont {Geddes}, \citenamefont
  {Benedetti},\ and\ \citenamefont {Leemans}}]{TeV}%
  \BibitemOpen
  \bibfield  {author} {\bibinfo {author} {\bibfnamefont {C.~B.}\ \bibnamefont
  {Schroeder}}, \bibinfo {author} {\bibfnamefont {E.}~\bibnamefont {Esarey}},
  \bibinfo {author} {\bibfnamefont {C.~G.~R.}\ \bibnamefont {Geddes}}, \bibinfo
  {author} {\bibfnamefont {C.}~\bibnamefont {Benedetti}}, \ and\ \bibinfo
  {author} {\bibfnamefont {W.~P.}\ \bibnamefont {Leemans}},\ }\href {\doibase
  10.1103/PhysRevSTAB.13.101301} {\bibfield  {journal} {\bibinfo  {journal}
  {Phys. Rev. ST Accel. Beams}\ }\textbf {\bibinfo {volume} {13}},\ p.\
  \bibinfo {pages} {101301} (\bibinfo {year} {2010})}\BibitemShut {NoStop}%
\bibitem [{\citenamefont {Stark}, \citenamefont {Toncian},\ and\ \citenamefont
  {Arefiev}(2016)}]{Stark}%
  \BibitemOpen
  \bibfield  {author} {\bibinfo {author} {\bibfnamefont {D.~J.}\ \bibnamefont
  {Stark}}, \bibinfo {author} {\bibfnamefont {T.}~\bibnamefont {Toncian}}, \
  and\ \bibinfo {author} {\bibfnamefont {A.~V.}\ \bibnamefont {Arefiev}},\
  }\href {\doibase 10.1103/PhysRevLett.116.185003} {\bibfield  {journal}
  {\bibinfo  {journal} {Phys. Rev. Lett.}\ }\textbf {\bibinfo {volume} {116}},\
  p.\ \bibinfo {pages} {185003}May (\bibinfo {year} {2016})}\BibitemShut
  {NoStop}%
\bibitem [{\citenamefont {Mora}\ and\ \citenamefont
  {T.M.~Antonsen}(1997)}]{Mora}%
  \BibitemOpen
  \bibfield  {author} {\bibinfo {author} {\bibfnamefont {P.}~\bibnamefont
  {Mora}}\ and\ \bibinfo {author} {\bibfnamefont {J.}~\bibnamefont
  {T.M.~Antonsen}},\ }\href@noop {} {\bibfield  {journal} {\bibinfo  {journal}
  {Phys. Plasmas}\ }\textbf {\bibinfo {volume} {4}},\ p.\ \bibinfo {pages}
  {217} (\bibinfo {year} {1997})}\BibitemShut {NoStop}%
\bibitem [{\citenamefont {Chen}\ \emph {et~al.}(2010)\citenamefont {Chen},
  \citenamefont {Sheng}, \citenamefont {Ma},\ and\ \citenamefont
  {Zhang}}]{chen_jap06}%
  \BibitemOpen
  \bibfield  {author} {\bibinfo {author} {\bibfnamefont {M.}~\bibnamefont
  {Chen}}, \bibinfo {author} {\bibfnamefont {Z.~M.}\ \bibnamefont {Sheng}},
  \bibinfo {author} {\bibfnamefont {Y.~Y.}\ \bibnamefont {Ma}}, \ and\ \bibinfo
  {author} {\bibfnamefont {J.}~\bibnamefont {Zhang}},\ }\href@noop {}
  {\bibfield  {journal} {\bibinfo  {journal} {J. Appl. Phys.}\ }\textbf
  {\bibinfo {volume} {99}},\ p.\ \bibinfo {pages} {025003} (\bibinfo {year}
  {2010})}\BibitemShut {NoStop}%
\bibitem [{\citenamefont {Lifschitz}\ and\ \citenamefont
  {Malka}(2020)}]{lifschitz_malka_njp12}%
  \BibitemOpen
  \bibfield  {author} {\bibinfo {author} {\bibfnamefont {A.~F.}\ \bibnamefont
  {Lifschitz}}\ and\ \bibinfo {author} {\bibfnamefont {V.}~\bibnamefont
  {Malka}},\ }\href@noop {} {\bibfield  {journal} {\bibinfo  {journal} {New J.
  Phys.}\ }\textbf {\bibinfo {volume} {14}},\ p.\ \bibinfo {pages} {053045}
  (\bibinfo {year} {2020})}\BibitemShut {NoStop}%
\bibitem [{\citenamefont {Nerush}\ and\ \citenamefont
  {Kostyukov}(2009)}]{kost_cep}%
  \BibitemOpen
  \bibfield  {author} {\bibinfo {author} {\bibfnamefont {E.~N.}\ \bibnamefont
  {Nerush}}\ and\ \bibinfo {author} {\bibfnamefont {I.~Y.}\ \bibnamefont
  {Kostyukov}},\ }\href {\doibase 10.1103/PhysRevLett.103.035001} {\bibfield
  {journal} {\bibinfo  {journal} {Phys. Rev. Lett.}\ }\textbf {\bibinfo
  {volume} {103}},\ p.\ \bibinfo {pages} {035001}Jul (\bibinfo {year}
  {2009})}\BibitemShut {NoStop}%
\bibitem [{\citenamefont {Ma}\ \emph {et~al.}(2016)\citenamefont {Ma},
  \citenamefont {Chen}, \citenamefont {Li}, \citenamefont {Yan}, \citenamefont
  {Huang}, \citenamefont {Chen}, \citenamefont {Sheng}, \citenamefont
  {Nakajima}, \citenamefont {Tajima},\ and\ \citenamefont
  {Zhang}}]{ma_screp16}%
  \BibitemOpen
  \bibfield  {author} {\bibinfo {author} {\bibfnamefont {Y.}~\bibnamefont
  {Ma}}, \bibinfo {author} {\bibfnamefont {L.}~\bibnamefont {Chen}}, \bibinfo
  {author} {\bibfnamefont {D.}~\bibnamefont {Li}}, \bibinfo {author}
  {\bibfnamefont {W.}~\bibnamefont {Yan}}, \bibinfo {author} {\bibfnamefont
  {K.}~\bibnamefont {Huang}}, \bibinfo {author} {\bibfnamefont
  {M.}~\bibnamefont {Chen}}, \bibinfo {author} {\bibfnamefont {Z.}~\bibnamefont
  {Sheng}}, \bibinfo {author} {\bibfnamefont {K.}~\bibnamefont {Nakajima}},
  \bibinfo {author} {\bibfnamefont {T.}~\bibnamefont {Tajima}}, \ and\ \bibinfo
  {author} {\bibfnamefont {J.}~\bibnamefont {Zhang}},\ }\href@noop {}
  {\bibfield  {journal} {\bibinfo  {journal} {Sc. Rep}\ }\textbf {\bibinfo
  {volume} {6}},\ p.\ \bibinfo {pages} {30491} (\bibinfo {year}
  {2016})}\BibitemShut {NoStop}%
\bibitem [{\citenamefont {Huijts}\ \emph {et~al.}(2020)\citenamefont {Huijts},
  \citenamefont {Andriyash}, \citenamefont {Rovige}, \citenamefont {Vernier},
  ,\ and\ \citenamefont {Faure}}]{CEP_observable}%
  \BibitemOpen
  \bibfield  {author} {\bibinfo {author} {\bibfnamefont {J.}~\bibnamefont
  {Huijts}}, \bibinfo {author} {\bibfnamefont {I.}~\bibnamefont {Andriyash}},
  \bibinfo {author} {\bibfnamefont {L.}~\bibnamefont {Rovige}}, \bibinfo
  {author} {\bibfnamefont {A.}~\bibnamefont {Vernier}}, , \ and\ \bibinfo
  {author} {\bibfnamefont {J.}~\bibnamefont {Faure}},\ }\href@noop {}
  {\bibfield  {journal} {\bibinfo  {journal} {arXiv}\ }\textbf {\bibinfo
  {volume} {2006}},\ p.\ \bibinfo {pages} {10566} (\bibinfo {year}
  {2020})}\BibitemShut {NoStop}%
\bibitem [{\citenamefont {Faure}\ \emph {et~al.}(2018)\citenamefont {Faure},
  \citenamefont {Gustas}, \citenamefont {Gu{\'{e}}not}, \citenamefont
  {Vernier}, \citenamefont {Böhle}, \citenamefont {Ouill{\'{e}}},
  \citenamefont {Haessler}, \citenamefont {Lopez-Martens},\ and\ \citenamefont
  {Lifschitz}}]{kHz_injection}%
  \BibitemOpen
  \bibfield  {author} {\bibinfo {author} {\bibfnamefont {J.}~\bibnamefont
  {Faure}}, \bibinfo {author} {\bibfnamefont {D.}~\bibnamefont {Gustas}},
  \bibinfo {author} {\bibfnamefont {D.}~\bibnamefont {Gu{\'{e}}not}}, \bibinfo
  {author} {\bibfnamefont {A.}~\bibnamefont {Vernier}}, \bibinfo {author}
  {\bibfnamefont {F.}~\bibnamefont {Böhle}}, \bibinfo {author} {\bibfnamefont
  {M.}~\bibnamefont {Ouill{\'{e}}}}, \bibinfo {author} {\bibfnamefont
  {S.}~\bibnamefont {Haessler}}, \bibinfo {author} {\bibfnamefont
  {R.}~\bibnamefont {Lopez-Martens}}, \ and\ \bibinfo {author} {\bibfnamefont
  {A.}~\bibnamefont {Lifschitz}},\ }\href@noop {} {\bibfield  {journal}
  {\bibinfo  {journal} {Plasma Phys.Contro Fusion}\ }\textbf {\bibinfo {volume}
  {61}},\ p.\ \bibinfo {pages} {014012} (\bibinfo {year} {2018})}\BibitemShut
  {NoStop}%
\bibitem [{\citenamefont {Chou}\ \emph {et~al.}(2013)\citenamefont {Chou},
  \citenamefont {Xu}, \citenamefont {Cardenas}, \citenamefont {Rivas},
  \citenamefont {Wittmann}, \citenamefont {Karsch},\ and\ \citenamefont
  {Veisz}}]{Veisz_2cycle}%
  \BibitemOpen
  \bibfield  {author} {\bibinfo {author} {\bibfnamefont {S.}~\bibnamefont
  {Chou}}, \bibinfo {author} {\bibfnamefont {J.}~\bibnamefont {Xu}}, \bibinfo
  {author} {\bibfnamefont {D.}~\bibnamefont {Cardenas}}, \bibinfo {author}
  {\bibfnamefont {D.}~\bibnamefont {Rivas}}, \bibinfo {author} {\bibfnamefont
  {T.}~\bibnamefont {Wittmann}}, \bibinfo {author} {\bibfnamefont {F.~K.~S.}\
  \bibnamefont {Karsch}}, \ and\ \bibinfo {author} {\bibfnamefont
  {L.}~\bibnamefont {Veisz}},\ }\href@noop {} {\bibfield  {journal} {\bibinfo
  {journal} {2013 Conference on Lasers and Electro-Optics - International
  Quantum Electronics Conference}\ } (\bibinfo {year} {2013})}\BibitemShut
  {NoStop}%
\bibitem [{\citenamefont {Sprangle}\ \emph {et~al.}(1992)\citenamefont
  {Sprangle}, \citenamefont {Esarey}, \citenamefont {Krall},\ and\
  \citenamefont {Joyce}}]{Sprangle_guide}%
  \BibitemOpen
  \bibfield  {author} {\bibinfo {author} {\bibfnamefont {P.}~\bibnamefont
  {Sprangle}}, \bibinfo {author} {\bibfnamefont {E.}~\bibnamefont {Esarey}},
  \bibinfo {author} {\bibfnamefont {J.}~\bibnamefont {Krall}}, \ and\ \bibinfo
  {author} {\bibfnamefont {G.}~\bibnamefont {Joyce}},\ }\href {\doibase
  10.1103/PhysRevLett.69.2200} {\bibfield  {journal} {\bibinfo  {journal}
  {Phys. Rev. Lett.}\ }\textbf {\bibinfo {volume} {69}},\ \unskip\ \bibinfo
  {pages} {2200--2203}Oct (\bibinfo {year} {1992})}\BibitemShut {NoStop}%
\bibitem [{\citenamefont {Kostyukov}, \citenamefont {Pukhov},\ and\
  \citenamefont {Kiselev}(2004)}]{kostyukov_pop04}%
  \BibitemOpen
  \bibfield  {author} {\bibinfo {author} {\bibfnamefont {I.}~\bibnamefont
  {Kostyukov}}, \bibinfo {author} {\bibfnamefont {A.}~\bibnamefont {Pukhov}}, \
  and\ \bibinfo {author} {\bibfnamefont {S.}~\bibnamefont {Kiselev}},\
  }\href@noop {} {\bibfield  {journal} {\bibinfo  {journal} {Phys. Plasmas}\
  }\textbf {\bibinfo {volume} {11}},\ p.\ \bibinfo {pages} {5256} (\bibinfo
  {year} {2004})}\BibitemShut {NoStop}%
\bibitem [{\citenamefont {Kalmykov}\ \emph {et~al.}(2009)\citenamefont
  {Kalmykov}, \citenamefont {Yi}, \citenamefont {Khudik},\ and\ \citenamefont
  {Shvets}}]{kalmykov}%
  \BibitemOpen
  \bibfield  {author} {\bibinfo {author} {\bibfnamefont {S.}~\bibnamefont
  {Kalmykov}}, \bibinfo {author} {\bibfnamefont {S.~A.}\ \bibnamefont {Yi}},
  \bibinfo {author} {\bibfnamefont {V.}~\bibnamefont {Khudik}}, \ and\ \bibinfo
  {author} {\bibfnamefont {G.}~\bibnamefont {Shvets}},\ }\href {\doibase
  10.1103/PhysRevLett.103.135004} {\bibfield  {journal} {\bibinfo  {journal}
  {Phys. Rev. Lett.}\ }\textbf {\bibinfo {volume} {103}},\ p.\ \bibinfo {pages}
  {135004}Sep (\bibinfo {year} {2009})}\BibitemShut {NoStop}%
\bibitem [{\citenamefont {Kostyukov}\ \emph {et~al.}(2009)\citenamefont
  {Kostyukov}, \citenamefont {Nerush}, \citenamefont {Pukhov},\ and\
  \citenamefont {Seredov}}]{kost_injection}%
  \BibitemOpen
  \bibfield  {author} {\bibinfo {author} {\bibfnamefont {I.}~\bibnamefont
  {Kostyukov}}, \bibinfo {author} {\bibfnamefont {E.}~\bibnamefont {Nerush}},
  \bibinfo {author} {\bibfnamefont {A.}~\bibnamefont {Pukhov}}, \ and\ \bibinfo
  {author} {\bibfnamefont {V.}~\bibnamefont {Seredov}},\ }\href {\doibase
  10.1103/PhysRevLett.103.175003} {\bibfield  {journal} {\bibinfo  {journal}
  {Phys. Rev. Lett.}\ }\textbf {\bibinfo {volume} {103}},\ p.\ \bibinfo {pages}
  {175003}Oct (\bibinfo {year} {2009})}\BibitemShut {NoStop}%
\bibitem [{\citenamefont {Yi}\ \emph {et~al.}(2010)\citenamefont {Yi},
  \citenamefont {Khudik}, \citenamefont {Kalmykov},\ and\ \citenamefont
  {Shvets}}]{austin_injection}%
  \BibitemOpen
  \bibfield  {author} {\bibinfo {author} {\bibfnamefont {S.~A.}\ \bibnamefont
  {Yi}}, \bibinfo {author} {\bibfnamefont {V.}~\bibnamefont {Khudik}}, \bibinfo
  {author} {\bibfnamefont {S.~Y.}\ \bibnamefont {Kalmykov}}, \ and\ \bibinfo
  {author} {\bibfnamefont {G.}~\bibnamefont {Shvets}},\ }\href@noop {}
  {\bibfield  {journal} {\bibinfo  {journal} {Plasma Phys.Contro Fusion}\
  }\textbf {\bibinfo {volume} {53}},\ p.\ \bibinfo {pages} {014012} (\bibinfo
  {year} {2010})}\BibitemShut {NoStop}%
\bibitem [{\citenamefont {Pukhov}(1999)}]{Pukhov_code}%
  \BibitemOpen
  \bibfield  {author} {\bibinfo {author} {\bibfnamefont {A.}~\bibnamefont
  {Pukhov}},\ }\href@noop {} {\bibfield  {journal} {\bibinfo  {journal} {J.
  Plasma Phys.}\ }\textbf {\bibinfo {volume} {61}},\ p.\ \bibinfo {pages} {425}
  (\bibinfo {year} {1999})}\BibitemShut {NoStop}%
\end{thebibliography}%

\end{document}